\documentclass[english]{article}
\usepackage[T1]{fontenc}
\usepackage[latin9]{inputenc}
\usepackage{color}
\usepackage{textcomp}
\usepackage{amstext}
\usepackage{amssymb}
\usepackage{esint}

\makeatletter

\newcommand{\lyxmathsym}[1]{\ifmmode\begingroup\def\b@ld{bold}
  \text{\ifx\math@version\b@ld\bfseries\fi#1}\endgroup\else#1\fi}

\newcommand{\lyxaddress}[1]{
\par {\raggedright #1
\vspace{1.4em}
\noindent\par}
}

\makeatother

\usepackage{babel}
\begin{document}

\title{\textbf{Effective state, Hawking radiation and quasi-normal modes
for Kerr black holes}}

\author{\textbf{$\ensuremath{^{1*}}$C. Corda, $^{2,**}$S. H. Hendi, $^{3,+}$R.
Katebi, $^{4,++}$ N. O. Schmidt}}

\maketitle

\lyxaddress{\begin{center}
${\ensuremath{^{1}}}$Institute for Theoretical Physics and Advanced
Mathematics Einstein-Galilei, Via Santa Gonda 14, 59100 Prato, Italy
\par\end{center}}

\lyxaddress{\begin{center}
${\ensuremath{^{2}}}$Physics Department and Biruni Observatory,
College of Sciences, Shiraz University, Shiraz 71454, Iran
\par\end{center}}

\lyxaddress{\begin{center}
${\ensuremath{^{3}}}$Physics Department, University of Massachusetts
Dartmouth, 285 Old Westport Road, North Dartmouth, MA 02747-2300,
USA
\par\end{center}}

\lyxaddress{\begin{center}
${\ensuremath{^{4}}\ensuremath{}}$Department of Mathematics, Boise
State University, 1910 University Drive, Boise, Idaho, 83725, USA
\par\end{center}}

\lyxaddress{\begin{center}
E-mail addresses: $\textcolor{blue}{\begin{array}{c}
 \ensuremath{^{*}}cordac.galilei@gmail.com,\ensuremath{^{**}}hendi@shirazu.ac.ir,\\
\ensuremath{^{+}}rkatebi.gravity@gmail.com,\ensuremath{^{++}}nathanschmidt@u.boisestate.edu 
\end{array}}$
\par\end{center}}
\begin{abstract}
The non-strictly continuous character of the Hawking radiation spectrum
generates a natural correspondence between Hawking radiation and black
hole (BH) quasi-normal modes (QNM). In this work, we generalize recent
results on this important issue to the framework of Kerr BHs (KBH).
We show that also for the KBH, QNMs can be naturally interpreted in
terms of quantum levels. Thus, the emission or absorption of a particle
is in turn interpreted in terms of a transition between two different
levels. At the end of the paper, we also generalize some concepts
concerning the ``effective state'' of a KBH. 
\end{abstract}

\section{Introduction}

The non-strictly thermal character \cite{key-1,key-2} of the Hawking
radiation spectrum \cite{key-3} shows that the emission of Hawking
quanta is also a non-strictly continuous process, by enabling a natural
correspondence between Hawking radiation and BH QNMs \cite{key-4}. 

Working with $G=c=k_{B}=\hbar=\frac{1}{4\pi\epsilon_{0}}=1$ (Planck
units), in a strictly thermal approximation the probability of emission
is \cite{key-1,key-2,key-3} 
\begin{equation}
\Gamma\sim\exp(-\frac{\omega}{T_{H}}),\label{eq: hawking probability}
\end{equation}
where $T_{H}\equiv\frac{1}{8\pi M}\,$ is the Hawking temperature
and $\omega$ is the energy-frequency of the emitted radiation.

By considering the important deviation from the strictly thermal character,
the correct probability of emission is indeed \cite{key-1,key-2}

\begin{equation}
\Gamma\sim\exp[-\frac{\omega}{T_{H}}(1-\frac{\omega}{2M})].\label{eq: Parikh Correction}
\end{equation}
The additional term $\frac{\omega}{2M}$ takes into due account the
conservation of energy, which arises from the fact that the BH contracts
during the process of radiation \cite{key-1,key-2}.

By introducing the \emph{effective temperature} \cite{key-4} 
\begin{equation}
T_{E}(\omega)\equiv\frac{2M}{2M-\omega}T_{H}=\frac{1}{4\pi(2M-\omega)}\label{eq: Corda Temperature}
\end{equation}
eq. ($\ref{eq: Parikh Correction}$) can be rewritten in Boltzmann-like
form \cite{key-4}

\begin{equation}
\Gamma\sim\exp[-\beta_{E}(\omega)\omega]=\exp(-\frac{\omega}{T_{E}(\omega)}),\label{eq: Corda Probability}
\end{equation}
where $\beta_{E}(\omega)\equiv\frac{1}{T_{E}(\omega)}$ and $\exp[-\beta_{E}(\omega)\omega]$
is the \emph{effective Boltzmann factor} appropriate for an object
with inverse effective temperature $T_{E}(\omega)\,$ \cite{key-4}.
The effective temperature represents the temperature of a strictly
thermal body that would emit the same total amount of radiation \cite{key-4}
and the ratio $\frac{T_{E}(\omega)}{T_{H}}=\frac{2M}{2M-\omega}$
represents the deviation of the radiation spectrum of a BH from the
strictly thermal feature \cite{key-4}. In other words, $M\,$ is
called the initial mass of the BH \emph{before} the emission and $M-\omega$
is the final mass of the BH \emph{after} the emission, where eqs.
($\ref{eq: Parikh Correction}$) and ($\ref{eq: Corda Temperature}$)
permit the BH \emph{effective mass} and \emph{effective horizon} definitions
\emph{during} the particle emission, i.e. during the BH's contraction
phase \cite{key-4} 
\begin{equation}
M_{E}(\omega)\equiv M-\frac{\omega}{2},\mbox{ }r_{E}(\omega)\equiv2M_{E}(\omega).\label{eq: effective mass}
\end{equation}
The \emph{effective quantities} introduced above are \emph{average
quantities} between the two states b\emph{efore and after the emission}
\cite{key-4}. $\emph{\ensuremath{T_{E}\:}}$is the inverse of the
average value of the inverses of the initial and final Hawking temperatures
\emph{before} the emission $T_{H\mbox{ initial}}=\frac{1}{8\pi M}$
and after the emission $T_{H\mbox{ final}}=\frac{1}{8\pi(M-\omega)}$,
respectively), while $\emph{\ensuremath{M_{E}\,}}$ is the average
of the initial and final masses, and $\emph{\ensuremath{r_{E}\,}}$
is the average of the initial and final horizons \cite{key-4}.

The interpretation of the particle emission is in terms of a \emph{quantum
transition} of frequency $\omega$ between the two discrete states
before and after the emission \cite{key-4}. From the tunneling point
of view, two separated classical turning points are joined by a trajectory
in imaginary or complex time when a tunneling happens \cite{key-1,key-4}.
As a consequence, the radiation spectrum is also discrete. The reason
is that, even if the statistical probability distribution ($\ref{eq: Parikh Correction}$)
and the statistical energy distribution are continuous functions at
a fixed Hawking temperature, such a Hawking temperature varies in
time with a discrete character in ($\ref{eq: Parikh Correction}$).
The size of the forbidden region that the tunneling particle traverses
is finite \cite{key-1} and this issue enables the introduction of
the effective temperature ($\ref{eq: Corda Temperature}$). Indeed,
in a strictly thermal approximation the turning points look to have
null separation \cite{key-1}. In that case, we do not know which
joining trajectory needs to be considered \cite{key-1}. In fact,
there is not any barrier \cite{key-1}. When the spectrum is instead
not strictly thermal the tunneling particle traverses a finite forbidden
region from $r_{initial}=2M$ to $r_{final}=2(M\lyxmathsym{\textminus}\omega),$
which works like a barrier \cite{key-1}. As a consequence, the Hawking
temperature and the energy emissions are also discrete quantities. 

We recall that the emitted energies are not only discrete, but also
countable. In fact, they have been counted in \cite{key-5,key-6},
where non-trivial correlations among radiations have been found in
energies governed by the spectrum ($\ref{eq: Parikh Correction}$).
The occurrence probability for a specific sequence of $n$ subsequent
energies $E_{i}=(E_{1},E_{2},....,E_{n})$ is \cite{key-5,key-6}

\begin{equation}
\Gamma(E_{1},E_{2},....,E_{n})=\Gamma(\sum_{1}^{n}E_{i}).\label{eq: formula cinese}
\end{equation}
If one considers two emissions with energies $E_{1}$ and $E_{2}$,
or one emission with energy $E_{1}+E_{2}$, the function %
\footnote{Notice that refs. \cite{key-23,key-24} are the first papers where
the Parikh-Wilczek method has been used in order to check if there
are correlations between emitted quanta. %
}

\begin{equation}
C\left[(E_{1}+E_{2}),E_{1},E_{2}\right]=\ln\Gamma(E_{1}+E_{2})-\ln\left[\Gamma(E_{1})\Gamma(E_{2})\right]=8\pi E_{1}E_{2}\label{eq: frequencies correlation}
\end{equation}
represents the statistical correlation between the emissions \cite{key-5,key-6}.

On the other hand, we recall that there are interesting proposals
on the non-strictly continuous character of Hawking radiation in the
literature \cite{key-7,key-8}. In general, quantum systems of finite
size are inclined to have a discrete energy spectrum instead of a
continuous one \cite{key-7}. In fact, the dynamics of a BH responsible
for the spectrum's character refer to both of the finite region enclosed
by the horizon \cite{key-7} and the finite size of the forbidden
region that the tunneling particle traverses \cite{key-1}. It is
exactly such a finite size which makes the process of tunneling to
be discrete instead of continuous \cite{key-4}. Hence, the BH energy
spectrum is discrete \cite{key-4,key-7,key-8}. 

The discrete character of the emission process and of the emission
spectrum implies a natural correspondence between Hawking radiation
and BH QNMs \cite{key-4}. Hence, QNMs can be naturally interpreted
in terms of quantum levels for both the emission and absorption of
particles \cite{key-4,key-9}.

By calling $l$ the angular momentum quantum number, the QNMs are
usually labeled as $\omega_{nl}.$ For each $l$ ($l\geq2$ for gravitational
perturbations), there is a second quantum number, namely the ``overtone''
one $n$ ($n=1,2,...$), which labels the countable sequence of QNMs
\cite{key-4,key-10,key-11,key-12}. For large $n$ the QNMs of the
Schwarzschild BH (SBH) become independent of $l,$ and, in a strictly
thermal approximation, have the following structure \cite{key-4,key-10,key-11,key-12}

\begin{equation}
\begin{array}{rl}
\omega_{n} & =\ln3\times T_{H}+2\pi i(n+\frac{1}{2})\times T_{H}+\mathcal{O}(n^{-\frac{1}{2}})\\
\\
 & =\frac{\ln3}{8\pi M}+\frac{2\pi i}{8\pi M}(n+\frac{1}{2})+\mathcal{O}(n^{-\frac{1}{2}}).
\end{array}\label{eq: quasinormal modes}
\end{equation}
Eq. ($\ref{eq: quasinormal modes}$) was originally obtained numerically
in \cite{key-13,key-14}. It was re-obtained through an analytic proof
in \cite{key-15,key-16}. 

The non-strictly thermal character of the BH spectrum permits us to
replace eq. ($\ref{eq: quasinormal modes}$) with \cite{key-4,key-9}
\begin{equation}
\begin{array}{rl}
\omega_{n} & =\ln3\times T_{E}(\omega_{n})+2\pi i(n+\frac{1}{2})\times T_{E}(\omega_{n})+\mathcal{O}(n^{-\frac{1}{2}})\\
\\
 & =\frac{\ln3}{4\pi\left[2M-(\omega_{0})_{n}\right]}+\frac{2\pi i}{4\pi\left[2M-(\omega_{0})_{n}\right]}(n+\frac{1}{2})+\mathcal{O}(n^{-\frac{1}{2}})
\end{array}\label{eq: quasinormal modes corrected}
\end{equation}
The Hawking temperature in eq. ($\ref{eq: quasinormal modes}$) has
been replaced by the effective temperature in eq. ($\ref{eq: quasinormal modes corrected}$)
\cite{key-4,key-9}. The physical interpretation is that the deviation
of the spectrum of BH QNMs from the strictly thermal feature implies
that the spacing of the poles in ($\ref{eq: quasinormal modes corrected}$)
coincides with the spacing $2\pi iT_{E}(\omega)$ expected for a \emph{non-thermal}
Green's function as a dependence on the frequency is present, while
in eq. ($\ref{eq: quasinormal modes}$) the spacing of the poles coincides
with the spacing $2\pi iT_{H}$ for a \emph{thermal} Green's function,
see \cite{key-4,key-9} for details. 

The physical solutions for the absolute values of the frequencies
in eq. ($\ref{eq: quasinormal modes corrected}$) is \cite{key-4,key-9}

\begin{equation}
(\omega_{0})_{n}=M-\sqrt{M^{2}-\frac{1}{4\pi}\sqrt{(\ln3)^{2}+4\pi^{2}(n+\frac{1}{2})^{2}}}\label{eq: radice fisica}
\end{equation}
In \cite{key-4,key-9} this solutions has been used to analyze important
properties and quantities of the SBH, like the horizon area quantization,
the area quanta number, the Bekenstein-Hawking entropy, its sub-leading
corrections and the number of micro-states, i.e. quantities which
are considered fundamental to realize the underlying unitary quantum
gravity theory. In this work, we generalize the SBH results in \cite{key-4,key-9}
to the framework of KBHs.

\section{Quasi-normal modes in Kerr black holes}

Following \cite{key-17}, the strictly thermal approximation of KBH
QNMs is given by \cite{key-17,key-18,key-19,key-20}

\begin{equation}
\omega(n)=\tilde{\omega}_{0}-i\left[4\pi T_{0}\left(n+\frac{1}{2}\right)\right],\label{eq: quasi-normal Kerr}
\end{equation}
where $\tilde{\omega}_{0}$ is a function of the BH parameters \cite{key-17}.
Although $T_{0}$ is called ``effective temperature'' in \cite{key-17},
it is \textbf{not} the same concept of effective temperature introduced
in \cite{key-4,key-9} that we discuss in the present paper. Indeed,
$T_{0}$ is a quantity introduced in \cite{key-21} within the framework
of Boltzmann weights and resonances. The two concepts must not be
confused. 

Calling $J$ the angular momentum of the BH and assuming

\begin{equation}
M^{2}\gg J\label{eq: molto maggiore}
\end{equation}
 gives 

\begin{equation}
T_{0}(J)\approx-\frac{T_{H}(J=0)}{2},\label{eq: T zero}
\end{equation}
where $T_{H}(J=0)$ is the Hawking temperature of the SBH. If one
wants to go beyond the strictly thermal approximation, then the replacement
$T_{H}\rightarrow T_{E}\,$ is needed, as $T_{E}$ (instead of $T_{H}$)
is the quantity associated to the emitted particle, i.e. the inverse
of the average value of the inverses of the initial and final Hawking
temperatures (\emph{before} the emission and \emph{after} the emission,
respectively). Hence, eq. ($\ref{eq: T zero}$) becomes 
\begin{equation}
T_{0}(J)\approx-\frac{T_{E}(J=0)}{2},\label{eq: T zero effective}
\end{equation}
 where $T_{E}(J=0)$ is the effective temperature of the SBH given
by eq. ($\ref{eq: Corda Temperature}$). 

As we are interested in highly excited BHs, i.e. $n$ is large, the
imaginary part of eq. ($\ref{eq: quasi-normal Kerr}$) becomes dominant.
Thus, setting $(\omega_{0})_{n}\equiv|\omega(n)|$, by using eqs.
($\ref{eq: quasi-normal Kerr}$) and ($\ref{eq: T zero effective}$)
we get immediately 

\begin{equation}
\triangle M_{n}=(\omega_{0})_{n-1}-(\omega_{0})_{n}=4\pi T_{0}=-2\pi T_{E}(J=0),\label{eq: variazione}
\end{equation}
for an emission involving the quantum levels $n$ and $n-1$.

The result ($\ref{eq: variazione}$) is totally consistent with the
results in \cite{key-4,key-9} for the SBH. In fact, in \cite{key-4,key-9}
we find 

\begin{equation}
\begin{array}{rl}
\triangle M_{n} & =(\omega_{0})_{n-1}-(\omega_{0})_{n}\\
\\
 & =\sqrt{M^{2}-\frac{1}{4\pi}\sqrt{(\ln3)^{2}+4\pi^{2}(n+\frac{1}{2})^{2}}}-\sqrt{M^{2}-\frac{1}{4\pi}\sqrt{(\ln3)^{2}+4\pi^{2}(n-\frac{1}{2})^{2}}},
\end{array}\label{eq: variazione 2}
\end{equation}
which for large $n$ becomes 
\begin{equation}
\begin{array}{c}
\triangle M_{n}\approx\sqrt{M^{2}-\frac{1}{2}(n+\frac{1}{2})}-\sqrt{M^{2}-\frac{1}{2}(n-\frac{1}{2})}.\end{array}\label{eq: variazione 3}
\end{equation}
On the other hand, by recalling that, as the BH's mass is decreasing
due to emissions of Hawking quanta, the BH's mass becomes \cite{key-4,key-9}

\begin{equation}
M_{n-1}\equiv\sqrt{M^{2}-\frac{1}{4\pi}\sqrt{(\ln3)^{2}+4\pi^{2}(n-\frac{1}{2})^{2}}}\label{eq: me 3}
\end{equation}
and 

\begin{equation}
M_{n}\equiv\sqrt{M^{2}-\frac{1}{4\pi}\sqrt{(\ln3)^{2}+4\pi^{2}(n+\frac{1}{2})^{2}}},\label{eq: me 4}
\end{equation}
at the levels $n-1$ and $n$, respectively. By using eq. ($\ref{eq: Corda Temperature}$),
we find that the BH's effective temperature for an emission involving
the quantum levels $n$ and $n-1$ is given by 
\begin{equation}
\begin{array}{rl}
T_{E}(\omega_{n}) & =\frac{1}{4\pi(2M-\omega_{n})}=\frac{1}{8\pi M_{E}(\omega_{n})}\\
\\
 & =\frac{1}{4\pi\left[\sqrt{M^{2}-\frac{1}{4\pi}\sqrt{(\ln3)^{2}+4\pi^{2}(n-\frac{1}{2})^{2}}}+\sqrt{M^{2}-\frac{1}{4\pi}\sqrt{(\ln3)^{2}+4\pi^{2}(n+\frac{1}{2})^{2}}}\right]}.
\end{array}\label{eq: Corda Temperature e-1 to e}
\end{equation}
For large $n$ eq. ($\ref{eq: Corda Temperature e-1 to e}$) becomes
\begin{equation}
\begin{array}{c}
T_{E}(\omega_{n})\approx\frac{1}{4\pi\left[\sqrt{M^{2}-\frac{1}{2}(n+\frac{1}{2})}+\sqrt{M^{2}-\frac{1}{2}(n-\frac{1}{2})}\right]}.\end{array}\label{eq: Corda Temperature e-1 to e approx}
\end{equation}
Thus, by combining eqs. ($\ref{eq: variazione}$), ($\ref{eq: variazione 3}$)
and ($\ref{eq: Corda Temperature e-1 to e approx}$), we see that
the result ($\ref{eq: variazione}$) is consistent with the results
in \cite{key-4,key-8} for the SBH if

\begin{equation}
\sqrt{M^{2}-\frac{1}{2}(n+\frac{1}{2})}-\sqrt{M^{2}-\frac{1}{2}(n-\frac{1}{2})}=-\frac{\frac{1}{2}}{\sqrt{M^{2}-\frac{1}{2}(n+\frac{1}{2})}+\sqrt{M^{2}-\frac{1}{2}(n-\frac{1}{2})}}.\label{eq: consistence}
\end{equation}
By multiplying each side of eq. ($\ref{eq: consistence}$) by $\sqrt{M^{2}-\frac{1}{2}(n+\frac{1}{2})}+\sqrt{M^{2}-\frac{1}{2}(n-\frac{1}{2})}$
one easily obtains the identity $-\frac{1}{2}=-\frac{1}{2}.$

\section{Effective states of Kerr black holes}

The introduction of the BH's \emph{effective state} enables the establishment
of additional effective quantities that should be important in the
framework of BH physics. Here, for a KBH of original mass $M$, we
define the effective state after a transition with QNM frequency $\omega$.

Following \cite{key-17}, we define the KBH's \emph{effective angular
momentum} $J_{E}(\omega)\equiv M_{E}(\omega)\alpha_{E}(\omega),$
where $\alpha_{E}(\omega)$ is the KBH's \emph{effective specific
angular momentum}. Thus, the KBH's \emph{outer and inner effective
horizons} can be defined as 

\begin{equation}
\begin{array}{c}
r_{E+}(\omega)\equiv M_{E}(\omega)+\sqrt{M_{E}^{2}(\omega)-\alpha_{E}^{2}(\omega)}\\
\\
r_{E-}(\omega)\equiv M_{E}(\omega)-\sqrt{M_{E}^{2}(\omega)-\alpha_{E}^{2}(\omega)}.
\end{array}\label{eq: outer and inner}
\end{equation}
Eq. ($\ref{eq: outer and inner}$) generalizes the results of eq.
(2) in \cite{key-17} to the non-strictly thermal case. The two expressions
in eq. ($\ref{eq: outer and inner}$) are the roots of the KBH's \emph{effective
quantity}

\begin{equation}
\triangle_{E}(\omega)\equiv r^{2}-2M_{E}(\omega)r+\alpha_{E}^{2}(\omega).\label{eq: delta efficace}
\end{equation}
If we also define 
\begin{equation}
\Sigma_{E}(\omega)\equiv r^{2}+\alpha_{E}^{2}(\omega)\cos^{2}\theta,\label{eq: sigma efficace}
\end{equation}
then we can introduce the KBH's effective line element

\begin{equation}
\begin{array}{c}
\left(ds^{2}\right)_{E}\equiv-\left(1-\frac{2M_{E}(\omega)r}{\Sigma_{E}(\omega)}\right)dt^{2}-\frac{4M_{E}(\omega)\alpha_{E}(\omega)r\sin^{2}\theta}{\Sigma_{E}(\omega)}dtd\varphi+\frac{\Sigma_{E}(\omega)}{\triangle_{E}(\omega)}dr^{2}\\
\\
+\Sigma_{E}(\omega)d\theta^{2}+\left(r^{2}+\alpha_{E}^{2}(\omega)+2M_{E}(\omega)\alpha_{E}^{2}(\omega)r\sin^{2}\theta\right)\sin^{2}\theta d\varphi^{2}.
\end{array}\label{eq: effective Kerr line element}
\end{equation}
Eq. ($\ref{eq: effective Kerr line element}$) takes into due account
the dynamical geometry of the KBH which emits and/or absorbs particles.

The introduced effective quantities permit us to define the KBH's
\emph{effective angular velocity}

\begin{equation}
\begin{array}{c}
\Omega_{E}(\omega)\equiv\frac{\alpha_{E}(\omega)}{r_{E+}^{2}(\omega)+\alpha_{E}^{2}(\omega)}\\
\\
=\frac{J_{E}(\omega)}{2M_{E}(\omega)\left(M_{E}^{2}(\omega)+\sqrt{M_{E}^{4}(\omega)-J_{E}^{2}(\omega)}\right)}.
\end{array}\label{eq: effective angular velocity}
\end{equation}
Therefore, we can define the KBH's \emph{effective horizon area} 

\begin{equation}
A_{E}(\omega)\equiv4\pi\left(r_{E+}^{2}(\omega)+\alpha_{E}^{2}(\omega)\right)=8\pi\left(M_{E}^{2}(\omega)+\sqrt{M_{E}^{4}(\omega)-J_{E}^{2}(\omega)}\right),\label{eq: effective horizon area}
\end{equation}
which permits us to define the KBH's \emph{effective temperature} 

\begin{equation}
\begin{array}{c}
\left(T_{KBH}\right)_{E}(\omega)\equiv\frac{r_{E+}(\omega)-r_{E-}(\omega)}{A_{E}(\omega)}\\
\\
=\frac{\sqrt{M_{E}^{4}(\omega)-J_{E}^{2}(\omega)}}{4\pi M_{E}(\omega)\left(M_{E}^{2}(\omega)+\sqrt{M_{E}^{4}(\omega)-J_{E}^{2}(\omega)}\right)}.
\end{array}\label{eq: KBH effective temperature}
\end{equation}
Now, the adiabatically invariant integral is written as \cite{key-17,key-22}
\begin{equation}
I=\int\frac{dM-\Omega dJ}{\omega}.\label{vagenas-adiabatic-invariant-integral}
\end{equation}
So how do we adjust Vagenas' eq. ($\ref{vagenas-adiabatic-invariant-integral}$)
to integrate with our KBH effective scenario? To answer this question
we must re-write eq. ($\ref{vagenas-adiabatic-invariant-integral}$)
to establish an effective formula that accepts the transition frequency
$\omega$ as input. Then, the transition frequency given by eq. (\ref{eq: variazione})
permits to define the KBH's \emph{effective adiabatic invariant} as
\begin{equation}
\begin{array}{rl}
I_{E}(\omega) & \equiv\int\frac{dM_{E}(\omega)-\Omega_{E}(\omega)dJ_{E}(\omega)}{2\pi T_{E}(\omega)}\\
\\
 & =2\left(M_{E}^{2}(\omega)+\sqrt{M_{E}^{4}(\omega)-J_{E}^{2}(\omega)}\right)\\
\\
 & -2M_{E}^{2}(\omega)\log\left(M_{E}^{2}(\omega)+\sqrt{M_{E}^{4}(\omega)-J_{E}^{2}(\omega)}\right),
\end{array}\label{eq: effective adiabatic invariant}
\end{equation}
 which generalizes the results of eqs. (25-26) in \cite{key-17} to
the non-strictly thermal case. 

Using eq. ($\ref{eq: effective horizon area}$), we can also generalize
the result of eq. (27) in \cite{key-17} to the non-strictly thermal
case 

\begin{equation}
I_{E}(\omega)=\frac{A_{E}(\omega)}{4\pi}-2M_{E}^{2}(\omega)\log\left(\frac{A_{E}(\omega)}{8\pi}\right).\label{eq: effective adiabatic invariant 2}
\end{equation}
Let us consider a KBH of original mass $M$ with the assumption (\ref{eq: molto maggiore}).
After a high number of emissions (and potential absorptions as the
BH can capture neighboring particles), the mass of the BH changes
from $M\;$ to the quantity $M_{n-1}$ of eq. (\ref{eq: me 3}) \cite{key-9}.
In the transition from the state with $n-1$ to the state with $n\;$
the mass of the black hole changes again from $M_{n-1}$ to the quantity
$M_{n}$ of eq. (\ref{eq: me 4}) \cite{key-9}. Now, the BH is excited
at a level $n$. We define the effective state for an emission from
the level $n-1$ to the level $n,$ with emission frequency $\triangle M_{n}$.
Therefore, the BH's effective mass is defined as 
\begin{equation}
M_{E}(\triangle M_{n})\equiv\frac{M_{n-1}+M_{n}}{2}=\frac{2M_{n}-\triangle M_{n}}{2}=M_{n}-\frac{\triangle M_{n}}{2}\label{noncontemp-effective-mass}
\end{equation}
where the BH's effective horizon is defined as 
\begin{equation}
r_{E}(\triangle M_{n})\equiv2M_{E}(\triangle M_{n}).\label{noncontemp-effective-horizon}
\end{equation}
Clearly, an absorption from the level $n$ to the level $n-1$ is
now potentially possible. In that case, the BH's effective mass and
the BH's effective horizon are the same.

Second, the KBH's effective angular momentum components are defined
as 
\begin{equation}
\alpha_{E}(\triangle M_{n})\equiv\frac{J_{E}(\triangle M_{n})}{M_{E}(\triangle M_{n})}\label{noncontemp-alpha-effective-angular-momentum}
\end{equation}
and using eqs. ($\ref{eq: delta efficace}-\ref{eq: sigma efficace}$)
to obtain 
\begin{equation}
\Delta_{E}(\triangle M_{n})\equiv r^{2}-2M_{E}(\triangle M_{n})r+\alpha_{E}^{2}(\triangle M_{n})\label{noncontemp-delta-effective-angular-momentum}
\end{equation}
and 
\begin{equation}
\Sigma_{E}(\triangle M_{n})\equiv r^{2}+\alpha_{E}^{2}(\triangle M_{n})\cos^{2}\theta.\label{noncontemp-sigma-effective-angular-momentum}
\end{equation}
Hence, eqs. ($\ref{eq: outer and inner}$), (\ref{noncontemp-effective-mass}),
and ($\ref{noncontemp-alpha-effective-angular-momentum}$) enable
us to construct the KBH's effective outer and inner horizons 
\begin{equation}
\begin{array}{rl}
r_{E+}(\triangle M_{n}) & \equiv M_{E}(\triangle M_{n})+\sqrt{M_{E}^{2}(\triangle M_{n})-\alpha_{E}^{2}(\triangle M_{n})}\\
\\
r_{E-}(\triangle M_{n}) & \equiv M_{E}(\triangle M_{n})-\sqrt{M_{E}^{2}(\triangle M_{n})-\alpha_{E}^{2}(\triangle M_{n})},
\end{array}\label{noncontemp-emit-outer-and-inner-horizons}
\end{equation}
which permit us to rewrite eq. (\ref{eq: effective Kerr line element})
as 
\begin{equation}
\begin{array}{c}
\left(ds^{2}\right)_{E}\equiv-\left(1-\frac{2M_{E}(\triangle M_{n})r}{\Sigma_{E}(\triangle M_{n})}\right)dt^{2}-\frac{4M_{E}(\triangle M_{n})\alpha_{E}(\triangle M_{n})r\sin^{2}\theta}{\Sigma_{E}(\triangle M_{n})}dtd\varphi+\frac{\Sigma_{E}(\triangle M_{n})}{\triangle_{E}(\triangle M_{n})}dr^{2}\\
\\
+\Sigma_{E}(\triangle M_{n})d\theta^{2}+\left(r^{2}+\alpha_{E}^{2}(\triangle M_{n})+2M_{E}(\triangle M_{n})\alpha_{E}^{2}(\omega)r\sin^{2}\theta\right)\sin^{2}\theta d\varphi^{2},
\end{array}\label{eq: effective Kerr line element 2}
\end{equation}
that takes into due account the dynamical geometry of the KBH which
emits or absorbs particles and the neighbouring quantum levels which
are involved in the transition.

Thus far, the introduced effective quantities authorize us to rewrite
the KBH's effective angular velocity in eq. (\ref{eq: effective angular velocity})
as 
\begin{equation}
\begin{array}{c}
\Omega_{E}(\triangle M_{n})\equiv\frac{\alpha_{E}(\triangle M_{n})}{r_{E+}^{2}(\triangle M_{n})+\alpha_{E}^{2}(\triangle M_{n})}\\
\\
=\frac{J_{E}(\triangle M_{n})}{2M_{E}(\triangle M_{n})\left(M_{E}^{2}(\triangle M_{n})+\sqrt{M_{E}^{4}(\triangle M_{n})-J_{E}^{2}(\triangle M_{n})}\right)},
\end{array}\label{noncontemp-effective-angular-velocity}
\end{equation}
to rewrite the KBH's effective horizon area in eq. ($\ref{eq: effective horizon area}$)
as 
\begin{equation}
\begin{array}{c}
A_{E}(\triangle M_{n})\equiv4\pi\left(r_{E+}^{2}(\triangle M_{n})+\alpha_{E}^{2}(\triangle M_{n})\right)\\
\\
=8\pi\left(M_{E}^{2}(\triangle M_{n})+\sqrt{M_{E}^{4}(\triangle M_{n})-J_{E}^{2}(\triangle M_{n})}\right)
\end{array}\label{noncontemp-effective-horizon-area}
\end{equation}
and to rewrite the KBH's effective temperature in eq. ($\ref{eq: KBH effective temperature}$)
as 
\begin{equation}
\begin{array}{c}
\left(T_{KBH}\right)_{E}(\triangle M_{n})\equiv\frac{r_{E+}(\triangle M_{n})-r_{E-}(\triangle M_{n})}{A_{E}(\triangle M_{n})}\\
\\
=\frac{\sqrt{M_{E}^{4}(\triangle M_{n})-J_{E}^{2}(\triangle M_{n})}}{4\pi M_{E}(\triangle M_{n})\left(M_{E}^{2}(\triangle M_{n})+\sqrt{M_{E}^{4}(\triangle M_{n})-J_{E}^{2}(\triangle M_{n})}\right)}.
\end{array}\label{noncontemp-effective-temperature}
\end{equation}
The KBH's effective adiabatic invariant in eq. ($\ref{eq: effective adiabatic invariant 2}$)
can be rewritten as 
\begin{equation}
I_{E}(\triangle M_{n})\equiv\frac{A_{E}(\triangle M_{n})}{4\pi}-2M_{E}^{2}(\triangle M_{n})\log\left(\frac{A_{E}(\triangle M_{n})}{8\pi}\right).\label{noncontemp-effective-adabatic-invariant}
\end{equation}
Considering Eq. (\ref{noncontemp-effective-horizon-area}), one can
show that 
\begin{equation}
\Delta A_{E}(\triangle M_{n})=16\pi M_{E}(\triangle M_{n})\left[1+\left(1-\frac{J_{E}^{2}(\triangle M_{n})}{M_{E}^{4}(\triangle M_{n})}\right)^{-\frac{1}{2}}\right]\triangle M_{n},\label{eq: mostrare}
\end{equation}
and therefore the KBH\textquoteright{}s \emph{effective area quanta
number} is defined as 
\begin{equation}
N_{E}(\triangle M_{n})\equiv\frac{A_{E}(\triangle M_{n})}{|\Delta A_{E}(\triangle M_{n})|}=\frac{M_{E}(\triangle M_{n})}{2\triangle M_{n}}\sqrt{1-\frac{J_{E}^{2}(\triangle M_{n})}{M_{E}^{4}(\triangle M_{n})}},\label{noncontemp-effective-area-quanta-number}
\end{equation}
which let us identify the KBH's \emph{effective Bekenstein-Hawking
entropy} as
\begin{equation}
\begin{array}{c}
(S_{KBH})_{BH}(\triangle M_{n})\equiv\frac{A_{E}(\triangle M_{n})}{4}\\
\\
=4\pi N_{E}(\triangle M_{n})M_{E}(\triangle M_{n})\left[1+\left(1-\frac{J_{E}^{2}(\triangle M_{n})}{M_{E}^{4}(\triangle M_{n})}\right)^{-\frac{1}{2}}\right]\cdot\triangle M_{n}.
\end{array}\label{noncontemp-effective-entropy}
\end{equation}
 As one can confirm, for $M_{E}^{2}\gg J_{E}$, the mentioned equations
reduce to those in \cite{key-4} for the SBH. 

We recall that, to the second order approximation, the BH's entropy
contains three parts: the usual Bekenstein-Hawking entropy, and two
sub-leading corrections, the logarithmic term and the inverse area
term \cite{key-4}

\begin{equation}
S_{total}=S_{BH}-\ln S_{BH}+\frac{3}{2A}.\label{eq: entropia totale}
\end{equation}
If one wants to satisfy the underlying quantum gravity theory, the
logarithmic and inverse area terms are requested \cite{key-4}. In
fact, for a better understanding of a BH's entropy in quantum gravity
it is imperative to go beyond Bekenstein-Hawking entropy and identify
the sub-leading corrections \cite{key-4}. Hence, the KBH's \emph{total
effective entropy} is \cite{key-4} 
\begin{equation}
S_{total}(\triangle M_{n})\equiv(S_{KBH})_{BH}(\triangle M_{n})-\ln(S_{KBH})_{BH}(\triangle M_{n})+\frac{3}{2A_{E}(\triangle M_{n})}.\label{noncontemp-total-effective-entropy}
\end{equation}
At this point, we have successfully defined the KBH's effective state.

We note that one can start with eq. ($\ref{noncontemp-effective-horizon-area}$)
and show that eq. ($\ref{noncontemp-effective-entropy}$) is valid
for $J_{E}\ll M_{E}^{2}$. In fact, for $J_{E}\ll M_{E}^{2}$ eq.
($\ref{noncontemp-effective-horizon-area}$) implies
\begin{equation}
A_{E}(\triangle M_{n})=16\pi M_{E}^{2}(\triangle M_{n})\label{eq: 49}
\end{equation}
Using the area law with Eq. (\ref{eq: 49}), one obtains 
\begin{equation}
(S_{KBH})_{BH}(\triangle M_{n})=4\pi M_{E}^{2}(\triangle M_{n}).
\end{equation}
 Using eq. ($\ref{noncontemp-effective-area-quanta-number}$), which,
for $J_{E}\ll M_{E}^{2}$, becomes 
\begin{equation}
N_{E}(\triangle M_{n})\equiv\frac{A_{E}(\triangle M_{n})}{|\Delta A_{E}(\triangle M_{n})|}=\frac{M_{E}(\triangle M_{n})}{2\triangle M_{n}},\label{becomes}
\end{equation}
one can replace one of $M_{E}(\triangle M_{n})$ in eq. (\ref{eq: 49})
in the following manner 
\begin{equation}
(S_{KBH})_{BH}(\triangle M_{n})=4\pi M_{E}(2N_{E}(\triangle M_{n})\triangle M_{n}),
\end{equation}
which is in agreement with eq. (46) when condition (12) is imposed
on eq. (46).

\section{Conclusion remarks}

In the first section of this paper, we briefly explained the important
issue that the non-strictly continuous character of the Hawking radiation
spectrum generates a natural correspondence between Hawking radiation
and BH QNMs \cite{key-4,key-9}. In doing so, we found that exemplifying
the discrete character of the BH energy spectrum, QNM transition process,
and radiation spectrum \cite{key-4,key-9} is essential to BH physics
because it authorizes us to encode this information as quantized structures
with well-defined effectives for mass, horizon, and temperature that
are fundamental to recognizing the underlying unitary quantum gravity
theory. 

Next, we took into due account the non-strictly thermal character
of the spectrum \cite{key-4,key-9}, which is also necessary to BH
physics because it permits us to use the effective quantities in \cite{key-4}
to generalize the SBH results in \cite{key-4,key-9} to the KBH framework.
In particular, we demonstrated in Section 2 that QNMs can be naturally
interpreted in terms of KBH quantum levels, where the obtained KBH
results are in full agreement with the SBH results in \cite{key-4,key-9}.
Therefore, these findings are meaningful because the effective quantities
in \cite{key-4,key-9} have been achieved for the stable four dimensional
SBH and KBH solutions in Einstein's general relativity.

In Section 3, we used the effective quantities in \cite{key-4,key-9}
as the foundation on which to construct the ``effective state''
of a KBH by generalizing the non-strictly thermal case results in
\cite{key-17}. It is imperative to express the KBH's effective state
because we need additional features and knowledge to consider in future
experiments and observations.

\section{Acknowledgements}

The authors thank an anonymous referee for useful criticisms and comments
which permitted to improve this paper.


\begin{thebibliography}{10}
\bibitem{key-1}M. K. Parikh, Gen. Rel. Grav. 36, 2419 (2004, First
Award at Gravity Research Foundation).

\bibitem{key-2}M. K. Parikh and F. Wilczek, Phys. Rev. Lett. 85,
5042 (2000).

\bibitem{key-3}S. W. Hawking, Commun. Math. Phys. 43, 199 (1975).

\bibitem{key-4}C. Corda, Int. Journ. Mod. Phys. D 21, 1242023 (2012,
Honorable Mention at Gravity Research Foundation); C. Corda, JHEP
1108, 101 (2011). 

\bibitem{key-5}B. Zhang, Q. Y. Cai, L. You and M. S. Zhan, Phys.
Lett. B 675, 98 (2009).

\bibitem{key-6}B. Zhang, Q. Y. Cai, M. S. Zhan and L. You, Annals
Phys. 326, 350 (2011). 

\bibitem{key-7}J. D. Bekenstein and V. F. Mukhanov, Phys. Lett. B
360, 7 (1995). 

\bibitem{key-8}Y. Yoon, arXiv:1210.8355 (2012). 

\bibitem{key-9}C. Corda, arXiv:1210.7747 (2012). 

\bibitem{key-10} S. Hod, Phys. Rev. Lett. 81 4293 (1998). 

\bibitem{key-11}S. Hod, Gen. Rel. Grav. 31, 1639 (1999, Fifth Award
at Gravity Research Foundation).

\bibitem{key-12}M. Maggiore, Phys. Rev. Lett. 100, 141301 (2008).

\bibitem{key-13}H. P. Nollert, Phys. Rev. D 47, 5253 (1993). 

\bibitem{key-14}N. Andersson, Class. Quant. Grav. 10, L61 (1993).

\bibitem{key-15} L. Motl, Adv. Theor. Math. Phys. 6, 1135 (2003). 

\bibitem{key-16} L. Motl and A. Neitzke, Adv. Theor. Math. Phys.
7, 307 (2003). 

\bibitem{key-17}E. C. Vagenas, JHEP 0811, 073 (2008). 

\bibitem{key-18}U. Keshet and S. Hod, Phys. Rev. D 76, 061501 (2007).

\bibitem{key-19}U. Keshet and A. Neitzke, Phys. Rev. D 78, 044006
(2008).

\bibitem{key-20}E. Berti, V. Cardoso and S. Yoshida, Phys. Rev. D
69, 124018 (2004).

\bibitem{key-21}U. Keshet and A. Neitzke, Phys. Rev. D 78, 044006
(2008).

\bibitem{key-22}A. J. M. Medved, Class. Quant. Grav. 25, 205014 (2008).

\bibitem{key-23}A. J. M. Medved and E. C. Vagenas, Mod. Phys. Lett.
A20, 1723 (2005).

\bibitem{key-24}M. Arzano, A. J. M. Medved and E. C. Vagenas, JHEP
0509, 037 (2005).\end{thebibliography}
\end{document}